\documentclass{aa}
\usepackage{graphics}
\begin{document}
\thesaurus{12(02.07.1; 11.03.01; 12.04.1; 13.25.2)} 
\title{Resolving the virial discrepancy in clusters of galaxies with
  modified Newtonian dynamics}
\author{R.H. Sanders}
\institute{Kapteyn Astronomical Institute, Groningen, The Netherlands}  
\date{Received     / Accepted   }
\titlerunning{The virial discrepancy in clusters of galaxies}

\maketitle

\begin{abstract}
A sample of 197 X-ray emitting clusters of galaxies is considered
in the context of Milgrom's modified Newtonian dynamics (MOND).
It is shown that the gas mass, extrapolated via an assumed $\beta$ model
to a fixed radius of 3 Mpc, is correlated with the gas temperature as 
predicted by MOND ($M_g \propto T^2$).  The observed temperatures are
generally consistent with the inferred mass of hot gas; no substantial
quantity of additional unseen matter is required in the context of MOND.
However, modified dynamics cannot resolve the strong lensing discrepancy
in those clusters where this phenomenon occurs.  The prediction is that
additional baryonic matter may be detected in the central regions of
rich clusters.


\end{abstract} 

\section{Introduction}

The discrepancy between the luminous mass
and the classical dynamical mass was first identified in clusters of
galaxies, the largest virialized systems in the Universe (Zwicky 1933).  
This discrepancy has been only partially alleviated by the subsequent 
detection of a substantial component of hot X-ray emitting intracluster
gas--  a component with a total mass which may, in rich clusters, exceed the 
stellar mass in galaxies by a factor of four or five (Jones \& Forman
1984, David et al. 1990, B\"ohringer et al. 1993).  Even considering
the contribution of this diffuse gas, the mass of detectable
matter fails by at least a factor of three-- more typically a factor of
10-- to account for the Newtonian dynamical mass of clusters.  
The traditional solution to this problem is to postulate the presence of
unseen matter which is most often assumed to be non-dissipative and 
non-baryonic.

Another solution lies in the possibility that there is no
substantial quantity of dark matter but that Newtonian
gravity or dynamics is not valid on the scale of large astronomical
systems.
Of several suggested alternatives to dark matter on extragalactic
scales, Milgrom's phenomenologically motivated modified 
Newtonian dynamics (MOND) is the most successful
in accounting for the systematics and details of the discrepancy in
galaxies.  The basic idea is that, below a critical acceleration,
$a_o\approx 10^{-8}$ cm/s$^2$, the magnitude of the gravitational
acceleration is given by $g=\sqrt{g_na_o}$
where $g_n$ is the Newtonian gravitational acceleration (Milgrom 1983a).
At higher accelerations, $g = g_n$ as usual.
The low acceleration limit directly implies the 
luminosity-velocity correlations observed
for galaxies ($L\propto v^4$)-- the Tully-Fisher relation for spirals 
and the Faber-Jackson relation for ellipticals--   as well as 
predicting that galaxy rotation curves become asymptotically
flat in the limit of large distance from the visible galaxy (Milgrom 1983b).

The success of the simple MOND prescription in predicting the detailed
shape of
the rotation curves of spiral galaxies from the observed distribution
of detectable matter is well-documented
(Begeman et al. 1991, Sanders 1996, McGaugh \& de Blok 1998a, b,
Sanders \& Verheijen 1998).
But MOND, as a modification of Newtonian gravity or inertia, 
must also account for the observed properties of
larger virialized systems which lie in the low-acceleration regime-- 
groups and clusters of galaxies.  Milgrom (1998)
has recently reconsidered small groups in the context of MOND 
and finds that the statistically averaged mass-to-light ratio in groups is 
on the order of unity, removing the necessity of dark matter (Milgrom 1998). 
I previously 
considered 20 X-ray emitting clusters (Sanders 1994, Paper 1),
and found that, 
for these objects, the mass predicted by MOND from the observed temperature
of the hot gas is consistent with the inferred mass of hot gas.
Moreover,  MOND predicts the observed gas mass-temperature
relation for clusters ($M_g\propto T^2$) which is, in effect, the high mass
continuation of the Faber-Jackson relation for elliptical galaxies.  

This original sample of X-ray emitting clusters was 
from the early analysis
of {\it Einstein} satellite data by Jones \& Forman (1984) with
temperature determinations by David et al. (1993).  Recently, 
{\it Einstein} data
for a much larger sample of 207 clusters has been compiled by 
White, Jones \& Forman (1997) who are interested primarily in the
properties of clusters with cooling flows.  The purpose of the present note 
is to consider this larger sample in the context of MOND.  I demonstrate
below that, with modified dynamics, the observed temperatures (or velocity
dispersions) of the clusters are consistent with the mass of hot gas 
estimated to be present within 1 to 3 Mpc of the cluster center.

\section{The sample:  gas mass and temperature determinations}

In their original sample, Jones \& Forman estimated
the gas mass inside 3 Mpc by fitting ``$\beta$ models'' to the observed
azimuthally averaged X-ray intensity distribution.  The $\beta$ model, 
which in most cases provides an adequate description of the mean run of X-ray 
intensity with projected radius, is 
characterized by a gas density distribution of the form
$$\rho = \rho_o\Bigl[1 + \Bigl({r\over {r_c}}\Bigr)^2\Bigr]^{-1.5\beta} 
\eqno(1)$$
where the core radius, $r_c$, the central density, $\rho_o$,
and the number, $\beta$ are parameters of the fit (see Sarazin 1988).  
The $\beta$ parameter varies from cluster to cluster 
but has a typical value on the order of 2/3 which implies that the
gas mass generally increases linearly with radius.  Obviously this 
must steepen at
some radius if the total gas mass is to converge, and it is clear that
assumptions about the outer radius are critical in assigning a gas mass
to a given cluster (so long as $\beta \le 1$).  Jones \& Forman estimated
the gas mass by extrapolating this model fit out to 3 Mpc even though
the actual parameters are usually determined by fitting over a smaller
region.

In their new compilation, White et al. (1997) use a different method to
estimate the gas mass-- the ``deprojection'' method invented by
Fabian et al. (1981).  Here one assumes, in addition to
perfect spherical symmetry, a model for the gravitational
potential of the cluster-- typically an isothermal sphere with a 
core radius and a velocity dispersion chosen to match that of the
observed velocity dispersion of the galaxies or the mean X-ray temperature.
One then assumes that the gas sits in hydrostatic equilibrium in this
model potential and adjusts the gas density and temperature distribution in
such a way as to reproduce the deprojected X-ray emissivity distribution.

The estimated mass of hot gas within a specific volume is fairly independent
of the method used (i.e., the $\beta$ model fitting or the deprojection 
method).  This is because the derived gas mass depends primarily upon the 
X-ray luminosity and the volume of the emission region (see White et al.
1994, White \& Fabian 1995).
However, with the deprojection method, the gas mass is
only determined within some radius, $r_{out}$, 
where the X-ray emission is lost in the background.
This cut-off radius varies from about 0.1 Mpc for the low-luminosity,
cooler clusters to 2 Mpc for the high-luminosity hot clusters.  
Therefore, the values of the gas mass tabulated by White et al. 
involve no uncertain extrapolation,
but it is evident that some fraction of the gas mass is missing-- a 
fraction which is probably greater in those clusters with smaller $r_{out}$,
i.e., the cooler clusters.

\section{Results}

Taking the numbers directly from Table 3 of White et al. (1997), who assume
a Hubble parameter of 50 km/s-Mpc, the
gas mass is plotted against temperature in Fig.\ 1.  
There are 197 clusters plotted here. The mass
of gas is that determined from the deprojection method inside radius
$r_{out}$ (column {\it viii}) and the temperature is the 
``reference'' temperature defined
from analysis of the X-ray spectrum for all of those clusters for which
this is available (column {\it vi} in parenthesis).  
This reference temperature rather than
the emission-weighted fitted temperature was used because this is 
comparable to that used in Paper 1 (the appearance of the plot is similar
if the fitted temperature is used instead).  
The error bars in mass and temperature are shown for those cases where
errors are given.  The open triangles are data for the 20 clusters from 
Paper 1, i.e., gas mass estimates from extrapolated $\beta$ models
(Jones \& Forman 1984).

\begin{figure}
  \resizebox{\hsize}{!}{\includegraphics{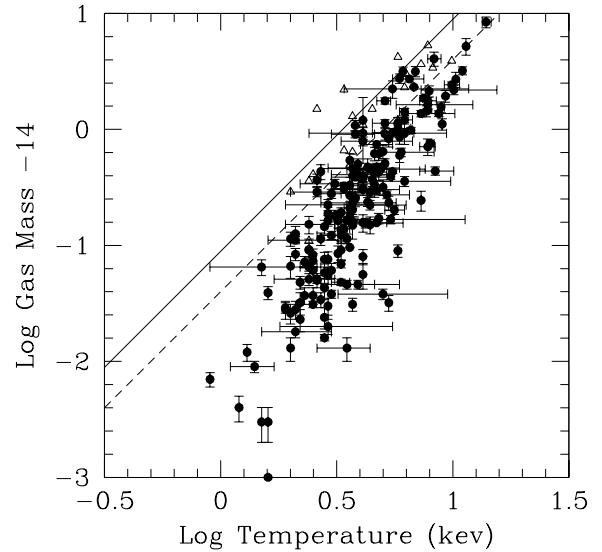}}
  \caption{A log-log plot of the mass of hot gas in units of
$10^{14}$ M$_\odot$ vs. gas temperature in kev 
for 197 X-ray emitting clusters from the tabulation by White et al. 1997.  
The gas mass is that determined by the deprojection method inside a
cutoff radius, $r_{out}$, which varies from cluster to cluster.  The 
temperature is the ``reference temperature'' as defined
by White et al.  The error bars are shown where given.  The MOND prediction
(eq.\ 3, $\beta$ = 2/3) is shown by the solid line and the extrapolated
Faber-Jackson relation by the dashed line.}
\end{figure}

These results are compared in Fig.\ 1 to the predictions of modified 
dynamics.
In the low acceleration limit of MOND, the spherically symmetric
equation of hydrostatic equilibrium (Milgrom 1984)  
may be immediately solved for the mass enclosed within radius r:
$$M_r  = {1\over{Ga_o}}{\Bigl({{kT_r}\over\mu}\Bigr)^2}{
\Bigl[{{d\,ln\rho}\over{dr}} + {d {ln\,T}\over{dr}}\Bigr]_r}^2 \eqno(2) $$
where k is the Boltzmann constant; $\mu$ is the mean molecular
weight; $a_o$ is the MOND acceleration parameter found to be
$0.8\times 10^{-8} h_{50}$ cm/s$^2$ from galaxy rotation curves (Begeman
et al. 1991); and the temperature and logarithmic gradients are given
at radius r. 
For a $\beta$ model, assumed to be isothermal, this may be written as
$$\Bigl({{M\over {10^{14}M_\odot}}}\Bigr) = 0.2 \beta^2 T^2 {h_{50}}^{-1}
\eqno(3)$$ with the temperature in kev (Sanders 1994).
This mass-temperature relation is shown by
the solid line in Fig.\ 1 where $\beta \approx 2/3$ is assumed. 

The high mass 
extrapolation of the the observed Faber-Jackson relation for elliptical
galaxies (Faber \& Jackson 1976), as calibrated by Fukugita \&
Turner (1991) scaled to $h_{50}=1$, is shown by the dashed line in Fig.\ 1, 
where the mass-to-light ratio for elliptical galaxies is assumed to be 7.5. 
This is shown here as well because, in the context of MOND, the gas 
mass-temperature relation for clusters is the high mass continuation
of the Faber-Jackson relation for ellipticals, apart from the anisotropy 
factor (Milgrom 1984) and the uncertainty of the mass-to-light ratio 
(Paper 1).

We see in Fig.\ 1 that the points from the new enlarged cluster
sample lie, except for the high temperature end, generally
below the MOND prediction for isothermal spheres and the extrapolated
Faber-Jackson relation for ellipticals and define a steeper relation
(a linear least-square fit gives a slope of 2.5).  The new mass estimates also 
lie below the those considered in Paper 1 which is not surprising because the 
estimates there were based upon extrapolations out to 3 Mpc. 
However, as noted above, the deprojection analysis misses a fraction
of the gas which is probably greater in those clusters with smaller $r_{out}$--
the lower temperature clusters.

There are two ways to correct for this effect.  The fairest way
is to consider only those systems in which $r_{out}$ is larger than some
specified value because this will eliminate clusters with the largest
fraction of missing gas.  This is shown in Fig.\ 2 where
only clusters with $r_{out} > 0.75$ Mpc (93 systems) are plotted.  
The agreement with the extrapolated Faber-Jackson law
is evident, but the masses are typically a factor of two or three below
the mass predicted by MOND for isothermal spheres with $\beta$ = 2/3. 

\begin{figure}
  \resizebox{\hsize}{!}{\includegraphics{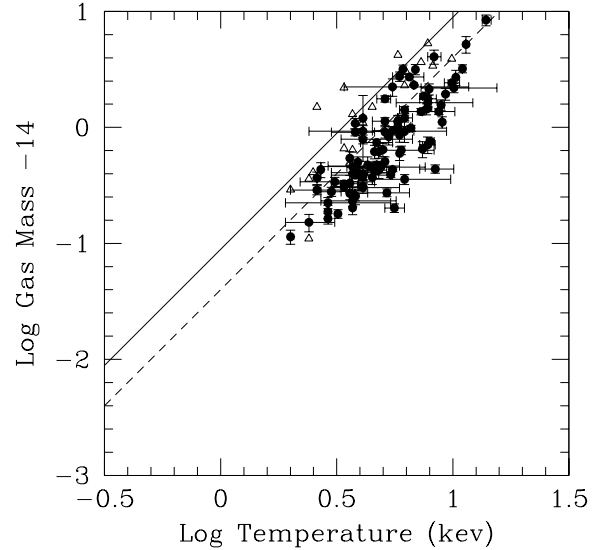}}
  \caption{Gas mass vs. temperature as in Fig.\ 1, but only 
for those cases (93 clusters) in which the cutoff radius for gas mass 
determination, $r_{out}$, exceeds 0.75 Mpc.}
\end{figure}

A second method of correcting for missing gas, less fair but
more consistent with the
extrapolated $\beta$ models, is to multiply the gas masses
given by White et al.\ by a factor of 3/r$_{\rm out}$ for each cluster.
Here one assumes that the
gas mass scales as radius ($\beta \approx 2/3$) at least out to 3 Mpc for
all clusters.  The results of this are shown in Fig.\ 3 where   
correlation between gas mass and temperature closely agrees in form and
amplitude with the MOND prediction of $M\propto T^2$.

\begin{figure}
  \resizebox{\hsize}{!}{\includegraphics{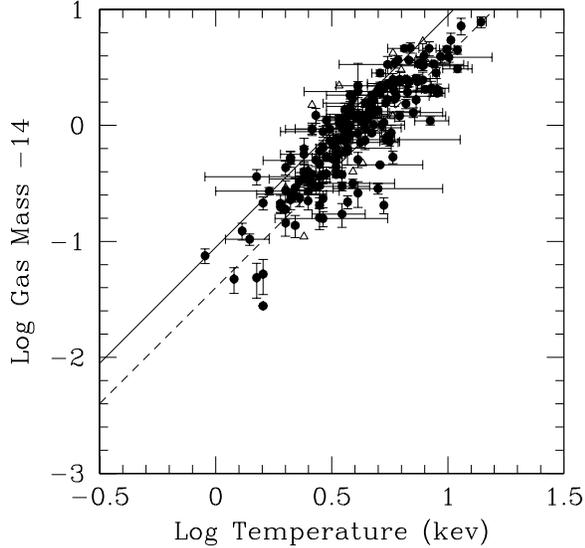}}
  \caption{Gas mass vs. temperature as in Fig.\ 1 but the
gas mass for all systems is scaled up by a factor of 3/r$_{out}$.
The assumption here is that the gas mass distribution is described by
a $\beta$ model with $\beta = 2/3$ which continues out to 3 Mpc for every
cluster.  This is a considerable extrapolation in several cases (a factor
of 30) and probably overestimates the total gas mass for some systems.}
\end{figure}

\section{Comments}

Considering the errors in temperature and in the estimated gas mass, the 
scatter in the observed gas mass-temperature relation is 
easily understandable.  To this should
be added the effect of the several idealized assumptions which are 
certainly not
realized in every case:  spherical symmetry, hydrostatic equilibrium,
isothermality, and thermal equilibrium between electrons and positive ions. 
Indeed, it is evident from eq.\ 2 that a pure $\beta$ model cannot be 
realized for an isothermal sphere in the low acceleration limit of 
modified dynamics-- the logarithmic density gradient must always 
steepen with radius.  Moreover, the contribution of the stellar mass
in galaxies to the total accounting of detectable matter cannot be
neglected in all cases (particularly in the lower temperature clusters).
Given these caveats, the approximate agreement of the extrapolated
gas mass with predictions of eq.\ 3 (Figs.\ 2 \& 3) demonstrates
that MOND can account for the overall virial
discrepancy in clusters within 1 to 3 Mpc.

However, there are actually two discrepancies in clusters of galaxies:  
that between the virial mass and the luminous mass and that between the
lensing mass and the luminous mass.  The discrepancy implied by weak lensing
due to clusters of galaxies (the slight systematic distortion of the
images of background galaxies) is generally consistent with the virial
discrepancy to within a factor of two or three (Wu \& Fang 1997, 
Allen 1998).  Therefore any relativistic
extention of MOND which preserves the relation between the weak field 
force and the deflection of light (i.e., $\theta = 2/c^2\int{g_\perp dl}$)
will also account for the lensing discrepancy (an example of such a theory
is given in Sanders, 1997).  

Some clusters also act as strong 
lenses; i.e., multiple images of background sources are formed by the
central regions of the clusters.  The critical surface density 
required for strong lensing is
$$\Sigma_c = {1\over {4\pi}} {{cH_o}\over {G}} F(z_l,z_s) \eqno(4)$$
where F is a dimensionless function of the lens and source redshifts
which depends upon the cosmological model (Blandford \& Narayan 1992); 
typically for clusters
and background sources at cosmological distances $F\approx 10$.
Modified dynamics applies in the limit of low 
accelerations or, equivalently, at surface densities below a value of
$\Sigma_M \approx a_o/G$ (Milgrom 1983a).  
Since, observationally it is found that $a_o \approx cH_o/6$, this implies 
that $\Sigma_c \approx 6\Sigma_M$;  that is to say, the critical surface
density for strong lensing is always greater than the upper limit
for MOND phenomenology.  {\it Strong lensing always occurs in
the Newtonian regime}.  Strong lensing observed in clusters typically
requires a total projected mass in the inner 100-200 kpc in 
excess of $10^{14} M_\odot$ which is evidently not present in the form
of hot gas.  Hitherto undetected matter does 
seem to be necessary in the cores of rich clusters which exhibit strong 
lensing, even with modified dynamics (see also Milgrom 1996).

This may be taken as a failure or as a prediction.  
That the tally of ordinary baryonic matter may not yet be complete is
suggested by the observations, in several clusters, of diffuse star light 
(Ferguson et al. 1998) and ultraviolet emission apparently from warm clouds 
(Mittaz et al. 1998).  Moreover, many X-ray clusters show evidence for 
cooling flows at some level (Sarazin 1988).  The gas cools, flows 
inward and disappears
into some, as yet, undetectable form.  The mass deposition rates are 
generally too low to be dynamically significant at the present epoch, 
but this may not have always been the case.

In summary, MOND goes a long way in resolving the virial discrepancy in
clusters.  It cannot resolve the strong lensing discrepancy in 
those clusters where this phenomenon is observed, but this  
leads to the prediction that more baryonic matter is present and
possibly detectable in the cores of rich clusters.  

I am grateful to M. Milgrom for useful comments on this work.

\end{document}